\documentstyle[aps,multicol,epsf,subfigure]{revtex}

\def\D{\partial}
\def\grad{\nabla}

\def\lap{\nabla^2}

\def\const{\mbox{const.}}
\def\eq{\begin{eqnarray}}
\def\qe{\end{eqnarray}}
\def\eqnn{\begin{eqnarray*}}
\def\qenn{\end{eqnarray*}}
\def\nn{\nonumber}
\def\be{\bm{e}}

\def\bn{\bm{n}}
\def\bq{\bm{q}}
\def\br{\bm{r}}

\def\simge{\;\lower3pt\hbox{$\stackrel{\textstyle >}{\sim}$}\;}
\def\simle{\;\lower3pt\hbox{$\stackrel{\textstyle <}{\sim}$}\;}
\def\bm#1{\mbox{\boldmath $#1$}}

\def\lrL#1{\left[#1\right]}

\def\lrS#1{\left(#1\right)}
\def\lrF#1{\left|#1\right|}
\def\lrA#1{\left\langle #1 \right\rangle}

\def\mycomment#1{}
\def\mycaption#1{\nopagebreak[4]\hspace{0mm}\\\begin{minipage}[h]{85mm}\caption{#1}\end{minipage}\vfill}

\def\f#1#2{\frac{#1}{#2}}
\def\der#1#2{\f{\D #1}{\D #2}}
\def\fder#1#2{\f{\delta #1}{\delta #2}}

\title  {Dynamics of orientational ordering in fluid membranes}
\author {Nariya Uchida\footnote{Present address: 
Department of Physics, Tohoku University, 980-8578, Japan.}
}
\address{Yukawa Institute for Theoretical Physics, 
Kyoto University, Kyoto 606-8502, Japan}
\date   {Submitted February 28, 2002; revised July 31, 2002}
\begin{document}
\draft

\bibliographystyle{prsty}

\maketitle
\begin{abstract}
We study the dynamics of orientational phase ordering
in fluid membranes.
Through numerical simulation we find
an unusually slow coarsening of topological texture,
which is limited by subdiffusive propagation of membrane curvature.
The growth of the orientational correlation length $\xi$
obeys a power law $\xi \propto t^w$ 
with $w < 1/4$ in the late stage.
We also discuss defect profiles and correlation patterns
in terms of long-range interaction mediated by curvature elasticity.
\end{abstract}
\pacs{
87.16.Dg, 
64.70.Md, 
64.60.Cn, 
05.70.Fh  
}

\begin{multicols}{2}


Fluid membranes exhibit a variety of shape transformations
due to their flexibility and sensitivity to changes in 
temperature and osmotic pressure.
They are most simply modeled by 
a structureless deformable surface with bending 
energy and a few constraints on its global geometry~\cite{Seifert}.
More diverse and prominent deformations
occur in membranes with internal degrees of freedom,
especially when they undergo a transition between 
different thermodynamic phases. 
An example is provided by a multi-component membrane 
in two-phase coexistence, deformation of which is controlled 
by a local coupling between composition and curvature.
The equilibrium morphology as well as the dynamics 
of phase separation have been addressed in a number 
of theoretical studies~\cite{Leibler,AKK,JL,Seifert2,TT,KR,KGL,JLS}, 
which have relevance to the so-called budding 
and other experimental observations.
\par
Another class of shape transformations
is characterized by a coupling to an
orientational degree of freedom.
It represents anisotropic in-plane shape 
and/or configuration of the constituent molecules.
The possibility of quasi-long-range 
orientational order has been proposed~\cite{HP,ML,FG}
and the resulting morphologies 
(such as tubules, sponges, and egg-cartons)
and defect energetics have been explored~\cite{ML,FG,LP,SSN,FP,OHSCM,SMS}.
A special case of this is the nematic order~\cite{FG,LP,SSN,FP,OHSCM},
the presence of which is suggested~\cite{FG} 
in the so-called ``gemini'' surfactants 
(each of which is 
a pair of molecules covalently linked at their polar heads)~\cite{Gemini}.
It is recently shown that 
the nematic order, in combination with chirality, 
accounts for the formation of helical ribbons 
by the material~\cite{OHSCM}.
\par
In this paper, we address the nematic phase ordering
from a dynamical perspective, which has been lacking so far.
Via numerical simulation, we show
that the coarsening of topological texture is 
greatly decelerated by the coupling 
to membrane deformation. 
The growth of texture size is asymptotically 
characterized by a power law with an exponent 
less than $1/4$, instead of the $1/2$ for ordinary 
nematic liquid crystals~\cite{CDTY,Toyoki,ZGG}.
We provide the interpretation of this novel 
non-equilibrium effect. To prepare for this, 
we also study how the curvature-orientation 
coupling controls defects and correlation 
structure in mechanical equilibrium.
Our results follow from a Ginzburg-Landau model in 
the simplest setting of a flat topology under a weak 
coupling condition.
It would be straightforward to include
more complicated though realistic aspects, such as
the stiffness anisotropy~\cite{HP,OHSCM}, 
non-flat topologies~\cite{LP}, and 
interlayer coupling~\cite{SSN,FP}.
%
%
%

We consider an initially flat membrane without any overhang,
and describe its profile in the Monge gauge as $z=h(\br)=h(x,y)$.
The orientational order parameter $Q_{ij}(\br)=Q_{ij}(x,y)$
is a symmetric traceless tensor
related to the scalar order parameter $S$ and 
nematic director $n_i$
as $Q_{ij} = S(n_i n_j - \delta_{ij}/2) (i,j=x,y)$.
The model free energy consists of three parts,
which are the homogeneous, Frank, 
and curvature-elastic contributions.
Under the assumption $|\grad h| \ll 1$, 
we retain the lowest order terms 
with respect to $\grad h$, as
\eq
F &=& \int d\br (f_{hom} + f_F + f_{curv}),
\label{Ftot}
\\
f_{hom} &=& \f{A}{2} Q_{ij}^2 + \f{C}{4} (Q_{ij}^2)^2,
\label{fL}
\\
f_F &=& \f{M}{2} \lrS{\D_i Q_{jk}}^2,
\label{fF}
\\
f_{curv} &=& \f{\kappa}{2} (\lap h)^2 + \alpha E_{ij} Q_{jk} \D_i \D_k h
\label{fcurv}
\qe
(summation over repeated indices is implied).
Here, $A$, $C$, and $M$ are the Landau-de Gennes coefficients~\cite{NB2},
$\kappa$ is the bending rigidity,
and $\alpha$ is the coupling constant.
Finally, $E_{ij}$ is the unit tensor for
an non-chiral nematic membrane~\cite{FG} and
the totally antisymmetric tensor
for a chiral nematic~\cite{OHSCM}.
From now on we concentrate on non-chiral membranes,
but discussions are parallel for chiral ones, 
as we shall see.
Neglecting long-range hydrodynamic interactions 
mediated by the solvent, 
we write the kinetic equations in the form
\eq
\der{Q_{ij}}{t} &=& - \Gamma_Q \lrS{\fder{F}{Q_{ij}}}^{(s)}
\label{dQdt}
\qe
where $(s)$ denotes the symmetric traceless part,
and 
\eq
\der{h}{t} 
= - \Gamma_h \fder{F}{h}
= - \Gamma_h \lrS{\kappa \lap \lap h + \alpha \D_i \D_j Q_{ij}}.
\label{dhdt}
\qe


To analyze the initial growth of order parameter 
upon a quench into the nematic phase, 
we linearize Eqs.(\ref{dQdt},\ref{dhdt})
with the aid of the unitary transformation
\eq
\left[\begin{array}{l}
Q_+(\bq) \\
Q_-(\bq) 
\end{array}\right]
&=&
\left[\begin{array}{rr}
  \cos 2 \zeta &  \sin 2 \zeta \\
- \sin 2 \zeta &  \cos 2 \zeta
\end{array}\right]
\left[\begin{array}{r}
Q_{xx}(\bq)
\\
Q_{xy}(\bq)
\end{array}\right],
\label{Qpm}
\\
\zeta &=& \arctan(q_y/q_x).
\qe
The three eigenmodes are expressed as
$$
e_{\pm}(\bq) = Q_+(\bq) + \f12 \lrL{c(q) 
\pm \sqrt{c(q)^2 + \f{4\Gamma_Q}{\Gamma_h}}} h(\bq)
$$
and 
$e_0(\bq) = Q_-(\bq)$,
where
$c(q) = -\kappa q^2/\alpha 
+ \Gamma_Q (A + M q^2)/(\Gamma_h \alpha q^2)$.
Their growth rates $\gamma_a = \D(\ln e_a)/{\D t}$ ($a=+,-,0$) read
\eq
\textstyle
\gamma_\pm(q) &=& \f12 \biggl[ -\Gamma_Q (A + Mq^2) - \Gamma_h \kappa q^4 
\nn\\
&& \qquad
\pm \Gamma_h \alpha q^2 \sqrt{c(q)^2 + \f{4\Gamma_Q}{\Gamma_h}}
\biggr], 
\qe
and $\gamma_0(q) = - \Gamma_Q (A + M q^2)$.
The spinodal point is located at $A = \alpha^2/\kappa$~\cite{FG,SSN}
as seen from the behavior of $\gamma_a(q)$ at small $q$.


The nonlinear dynamics of phase ordering 
is studied by numerically integrating 
Eqs.(\ref{dQdt},\ref{dhdt}) on a square lattice.
We choose $A=-1$, $C=20$, $M=1$, $\kappa=20$, and $\alpha=1$
as the standard parameter set,
with the mesh size $\Delta x = 1$.
The Landau coefficients give the equilibrium 
scalar order parameter  $S_{eq} \sim \sqrt{2|A|/C} \simeq 0.3$
and the defect core size $l_{core} \sim \sqrt{M/|A|} = 1$.
The large ratio $\kappa/M = 20$ is not unrealistic if we consider
the large stiffness $\kappa \sim 20 k_BT$ usually found in biological 
membranes.
We impose periodic boundary conditions 
on a $512\times512$ lattice.
For the initial condition, random numbers 
uniformly distributed in $[-0.1,0.1]$ are
assigned to each component of $Q_{ij}$.
The kinetic equations are
integrated using the Euler scheme
with $\Gamma_Q=0.1$, $\Gamma_h=0.1$ 
and the step increment $\Delta t=0.025$.
Note that the flat membrane approximation breaks down 
as the texture size exceeds the equilibrium 
curvature radius $\sim \kappa/(\alpha S_{eq})$.
All the data shown below are taken from the time region
in which $\lrA{|\grad h|} < 0.15$
(we denote a spatial average by $\lrA{\cdots}$).


Shown in Fig.1 are snapshots of
the Schlieren texture $Q_{xy}^2(\br)$
and the mean curvature $\lap h(\br)$.
In the late stage, the coarsening proceeds
via pair annihilation of topological defects with 
$s=+1/2$ and $-1/2$, where $s$ is the winding number
for the apolar vector $\bn$ (or the disclination strength~\cite{Chandrasekhar}).
It is seen that
a $-1/2$ defect accompanies three lobes of positive mean curvature,
while a $+1/2$ defect has one.
Where the director points to the defect core,
$\lap h$ is negative (positive) for a +1/2 (-1/2) defect (resp.).
To understand this, we note that, the membrane shape 
in the latest stage of coarsening is locally well equilibrated 
with respect to the order parameter configuration.
Then, for analysis of {\it static} correlation properties, 
we can assume the mechanical equilibrium condition 
$\delta F/\delta h = 0$, 
or 
\eq
\lap (\lap h) = -\f{\alpha}{\kappa} \D_i \D_j Q_{ij}.
\label{poisson}
\qe
This can be solved in the Fourier space as 
$h(\bq) = (\alpha/\kappa q^2) Q_+(\bq)$,
substitution of which into (\ref{fcurv})
gives the effective curvature elastic energy as
\eq
F_{curv}^{(e\!f\!f)} 
&=& -\f{\alpha^2}{2\kappa} \int \f{d\bq}{(2\pi)^2} 
\lrF{Q_+(\bq)}^2.
\label{Fcurv}
\qe
It becomes dominant 
over Frank elasticity when the texture size is larger than
the characteristic length,
\eq
\lambda = \sqrt{\kappa M}/\alpha.
\qe
Within this distance from a defect,
the optimum configuration $Q_{ij}(\br)$ 
is little affected by the elastic coupling
and approximately minimizes Frank elastic energy.
By further assuming $S=\const$, $\bn = (\cos \theta, \sin \theta)$,
and the polar coordinate system $(r, \phi)$,
the approximation gives~\cite{Chandrasekhar}
\eq
\theta = s (\phi - \phi_0), \quad \phi_0 = \const
\qe
for a defect of index $s$ located at the origin. 
The Poisson equation (\ref{poisson}) for mean curvature
has the special solution,
\eq
\f{\kappa}{\alpha S} \lap h 
=
\left\{
\begin{array}{ll}
\displaystyle{
\f{s}{2s-2}}\cos \lrL{2 (\theta - \phi)} & (s \neq 1),
\\
0   & (s=1).
\end{array}
\right.
\label{laph}
\qe
This qualitatively explains the mean curvature profile
obtained by simulation.
It also shows that a $+1/2$ defect carries much larger 
bending energy than its anti-defect.
On the other hand,
the anisotropic part of the curvature 
\eq
H_{ij} = \D_i \D_j h - (\lap h/2) \delta_{ij}
\qe
is proportional to $Q_{ij}$ in a uniform equilibrated system,
and the principal curvature axis is parallel to the director.
This holds also in the coarsening system
in the late stage and except in the vicinity of defects,
as we confirmed numerically. 
%


The long-range elastic interaction
also modifies orientational correlation 
at lengthscales larger than $\lambda$.
In the real space,
and under the approximation $S = \const$, 
Eq.(\ref{Fcurv}) 
is rewritten as 
\eqnn
F_{curv}^{(e\!f\!f)} &=& -\f{\alpha^2 S^2}{16 \pi \kappa}
\int \!\!\! \int d\br d\br'
\f{
\cos 2\lrL{\theta_{rel}(\br, \br') + \theta_{rel}(\br', \br)}
}{
\lrF{\br-\br'}^2
},
\qenn
where $\theta_{rel}(\br, \br')$ is the angle
between $\bn(\br)$ and $\br' - \br$~\cite{PL}.
It suggests that orientational correlation
is enhanced in directions parallel and 
perpendicular to the local director,
while it is suppressed in oblique directions.
This can be quantified by
the relative orientation correlation function,
which we define as
\eqnn
G_{rel}(\br - \br') &=& \lrA{
Q_{ij}(\br) Q_{ij}\lrS{\br+ U(\bn(\br)) \cdot (\br'-\br)}
}
\qenn
where $U(\bn(\br))$ is a matrix of rotation
that maps $\bn(\br)$ onto $\be_x$.
Simulation result in Fig.2 shows that
the expected tendency is prominent in 
the region $|\br-\br'| \gg \lambda$, 
where correlation in oblique directions is negative.


The coarsening kinetics is monitored through
the orientational correlation length $\xi_Q$,
which is defined by the correlation function
$G_Q(|\br-\br'|) = \lrA{Q_{ij}(\br)Q_{ij}(\br')}$
via
$G_Q(\xi_Q)/G_Q(0) = 1/2$.
Similarly, the curvature correlation length $\xi_H$
is defined as the half-value decay length of
$G_H(|\br-\br'|)=\lrA{H_{ij}(\br)H_{ij}(\br')}$.
In Fig.3, the orientational correlation length
is plotted as a function of time.
In the absence of coupling,
$\xi_Q$ is well fitted by a power law
$\xi_Q \propto t^{w_0}$ with $w_0 = 0.43 \pm 0.02$,
in good agreement with previous results~\cite{ZGG,BB}. 
Remarkably, the elastic coupling makes
the growth much slower.
The instantaneous growth exponent
\eq
w_Q(t)= d(\ln\xi_Q)/d(\ln t)
\label{exponent}
\qe
decreases from an initial value $\simeq w_0$,
and then converges to a much smaller value $w_{\infty}$
as $t \to \infty$.
Deviation from the initial value starts
earlier for a stronger coupling.
In the intermediate stage, $w_Q(t)$ shows
an undershoot and even turns negative
for most of the parameters we studied.
The asymptotic exponent $w_{\infty}$ is estimated
to be $0.12\pm0.02$ for $\alpha=2$
and $0.13\pm0.01$ for $\alpha=3$~\cite{NB1}.
On the other hand, the curvature coarsening 
exponent $w_H(t) = d(\ln\xi_H)/d(\ln t)$
first decreases monotonically 
and then converges to 
the same value $w_{\infty}$ within error bars.
In the bottom of the figure we plot the ratio $\xi_Q/\xi_H$,
which decays to unity after an initial increase.
In the same figure, we indicate the time at which
the ratio $\lrA{|f_{curv}|}/\lrA{f_F}$ reaches $2$.
It shows that the two lengths merge 
as the curvature-orientational coupling
becomes dominant over Frank elasticity.
\par
How does the coupling decelerate coarsening?
First we note that the effective elastic
interaction (\ref{Fcurv}) does not provide an explanation.
In the mechanical equilibrium,
the free energy density $f_{curv}$ is reduced
to its homogeneous minimum except in a region
of size $\sim \lambda$ around each defect,
in which Frank and curvature elastic energies
are balanced.
Thus the $\xi_Q$-dependence of $f_{curv}$ is
that of the defect density $\propto 1/\xi_Q^2$,
which is also the scaling of Frank elastic energy
(except for a logarithmic correction).
Therefore, the effective elastic interaction
cannot affect the coarsening exponent.
This was confirmed by an additional simulation
that directly implements the long-range interaction
using Fourier transformation.
It shows that
(i) coarsening curves $\xi_Q(t)$ and $\xi_H(t)$ are
both well fitted by the exponent $0.43$ for null coupling;
(ii) dissipation rates $d \lrA{f_F}/dt$ and $d \lrA{f_{curv}}/dt$ 
are of the same order
and their ratio is roughly constant in the late stage.
\par
Thus we can attribute the origin of slow coarsening
to the dynamics of shape change, Eq.(\ref{dhdt}).
It is subdiffusive and
the source term ($\propto \D_i \D_j Q_{ij}$)
is small except near defects.
According to a simple dimensional counting,
the mean defect separation scales like $\xi_Q$
and the timescale of curvature
relaxation is proportional to $\xi_Q^4$.
If $w_\infty > 1/4$, the coarsening of curvature field
cannot keep up with that of order parameter,
and hence $f_{curv}$ increases.
This is energetically disfavored
once the coupling becomes dominant over Frank elasticity.
Therefore, the asymptotic exponent cannot exceed $1/4$.
The reason why $w_\infty$ is even smaller than $1/4$
is not very simple.
A possible explanation is
the logarithmic correction to the
$\xi_Q$-dependence of Frank elastic energy.
In the XY model and ordinary nematic fluids in 2D,
it causes a logarithmic correction
to the power law~\cite{Bray} and
the apparent exponent is smaller than $1/2$
in finite time simulations~\cite{Toyoki,ZGG,BB,MG,YPKH}.
A similar effect is expected for the present case,
because the coarsening is still driven by
Frank elasticity.

\par
The non-monotonicity of $\xi_Q(t)$ in the intermediate stage 
corresponds to the fact that Frank elastic energy is temporally
sacrificed to reduce curvature elastic energy,
which has initially grown on the smaller scale $\xi_H$.
Because $\xi_Q$ grows faster than $\xi_H$
in the initial stage,
it is at least necessary that $w_Q < w_H$ holds temporally
for the two lengths to finally merge.


Some remarks are in order.
First, hydrodynamic flow of the solvent
effectively modifies the mobility as 
$\Gamma_h(q) = 1/(4 \eta q)$ ($\eta$: viscosity)~\cite{Seifert}.
However, the curvature dynamics is still subdiffusive, 
and the coarsening exponent cannot exceed $1/3$.
We have performed a simulation incorporating the hydrodynamic 
interaction, by solving Eq.(6) with $\Gamma_h(q)$ above
in the Fourier space and transforming back
each time step to solve Eq.(\ref{dQdt}) in the real space.
We found the coarsening exponent $\xi_Q$ to be $0.17 \pm 0.03$
(for $\alpha=2$ and $\eta=1.25$), which confirms the 
weakness of hydrodynamic accerelation. The details of
this simulation will be presented elsewhere.
\par
Second, the mechanism of slow coarsening
is unique to membranes with a continuous degree of freedom
and is irrelevant to phase separation dynamics,
which is described by a scalar order parameter~\cite{TT,KR,KGL}.
This has the following reason.
In phase separation, significant changes in
the order parameter occur only at domain boundaries.
However, domain boundaries are also the sources of 
curvature propagation, as we can see from the corresponding 
dynamic equation for $h$.
Therefore, the membrane shape can immediately follow 
the composition change.
In contrast, the orientational order parameter 
changes significantly even far from moving defects.
In order to adapt to this, shape deformations 
spread from the defects and over the whole texture, 
which is slow.


Finally we consider the effect of chirality.
With the antisymmetric coupling between curvature and $Q_{ij}$,
a straightforward calculation gives the elastic interaction as
\eq
F^{(e\!f\!f)}_{curv} &=&  -\f{\alpha^2}{2\kappa} \int \f{d\bq}{(2\pi)^2} 
\lrF{Q_-(\bq)}^2.
\label{Fcurv2}
\qe
The chirality reverses the correlation anisotropy so that 
the director correlation is enhanced in oblique directions.
However, the coarsening law is unaltered
since the last term in the membrane's dynamic equation (\ref{dhdt}) 
is still proportional to $\grad \grad {\sf Q}$ and sources of 
curvature propagation are localized around defects.
%

%
%

This work is supported 
by the Grant in Aid for
Scientific Research from Japan Society for
the Promotion of Science.
%
%

%
%
%
%
%
\begin{figure}[h]
\epsfxsize=250pt \epsffile{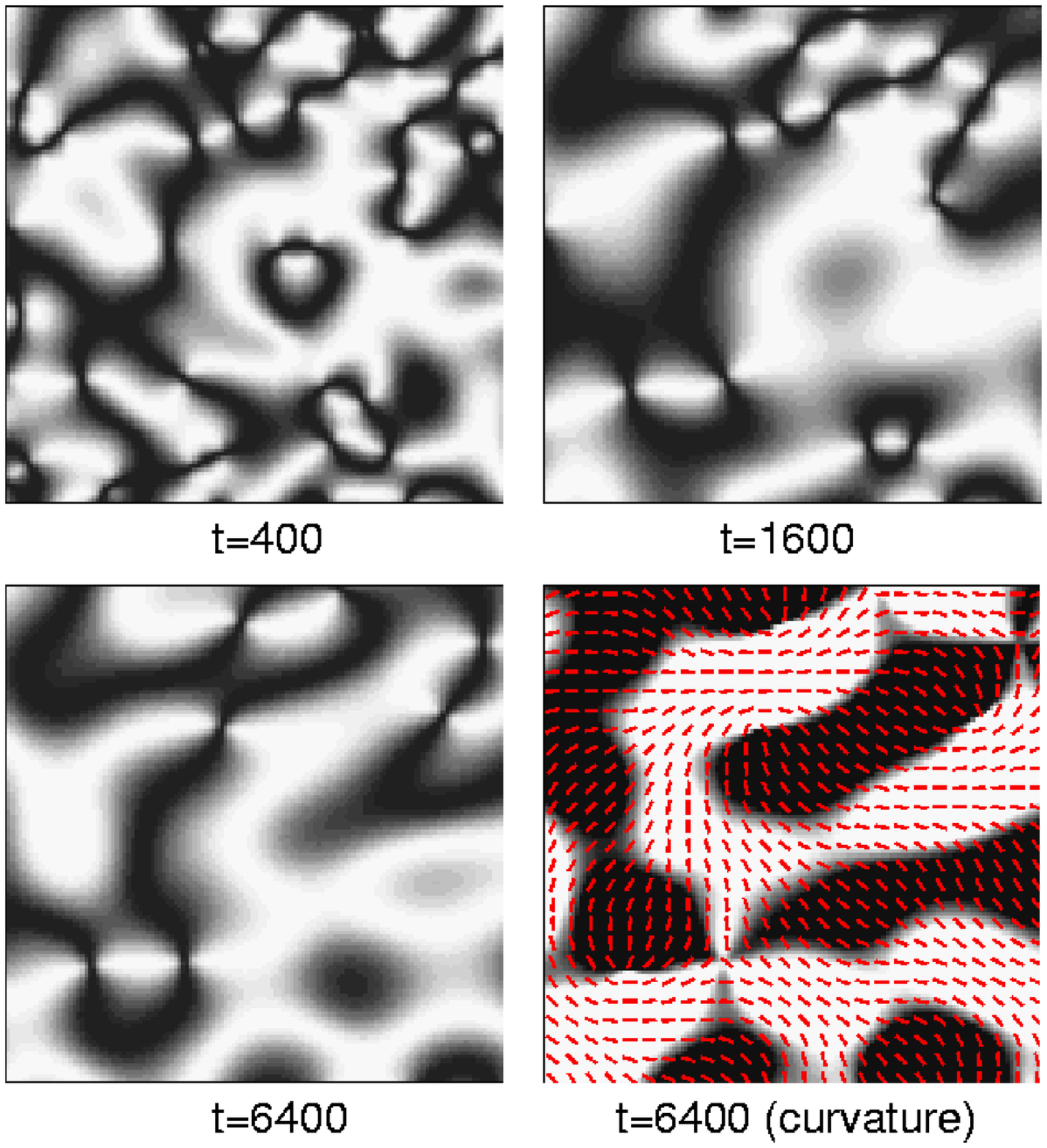}
\mycaption{
\label{fig1}
Snapshots of $Q_{xy}^2({\bf r})$ on a $128^2$
portion of the lattice, at $t=400$, $1600$, and $6400$.
The last panel shows the mean curvature $\lap h({\bf r})$ 
with the director field in short lines.
}
\end{figure}

%
%
\begin{figure}[h]
\epsfxsize=200pt \epsffile{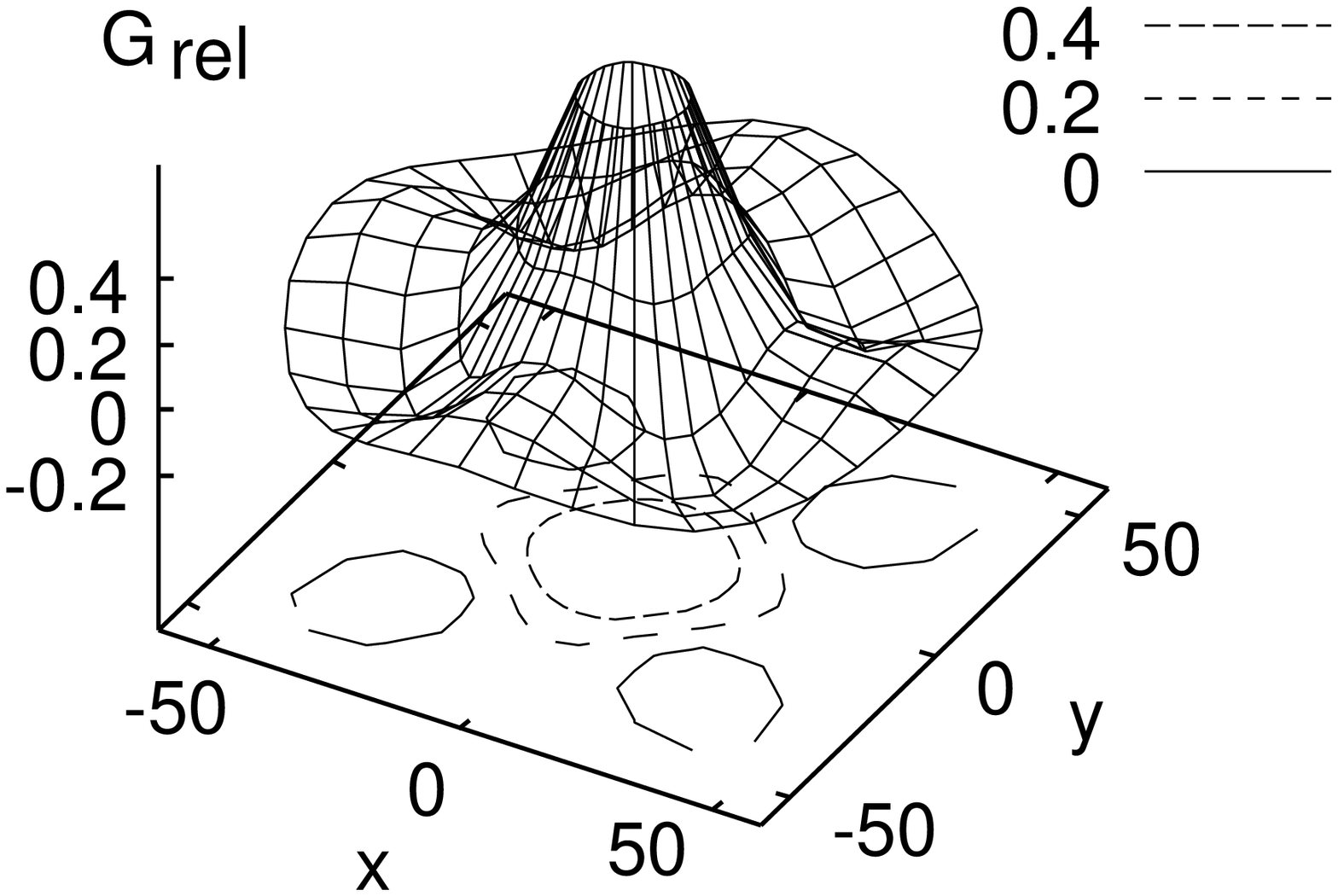}
\mycaption{
\label{fig2}
Relative orientation correlation function $G_{rel}({\bf r})$
at $t=6400$ (averaged over $10$ independent runs).
}
\end{figure}

%
%
\begin{figure}[h]
\epsfxsize=240pt \epsffile{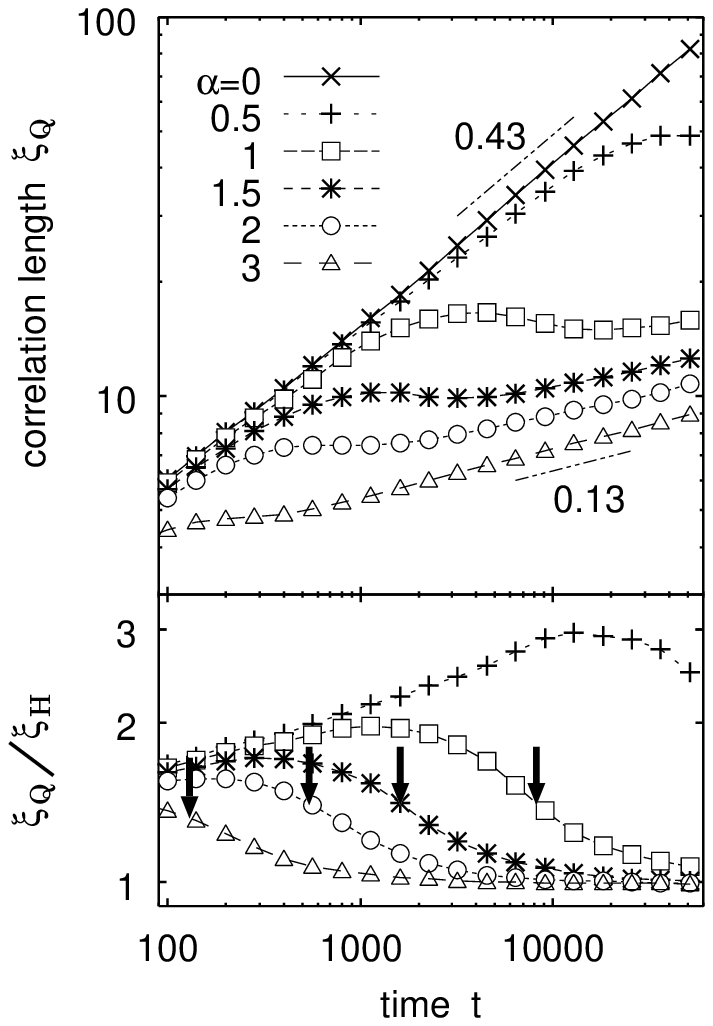}
\mycaption{
\label{fig3}
Top: orientational correlation length $\xi_Q$ versus time.
The lines corresponding to $w_Q=0.13$ and $0.43$ 
are drawn as guides to the eye.
Bottom: the ratio between orientational
and curvature correlation lengths.
Arrows indicate the time at which $\lrA{|f_{curv}|}/\lrA{f_F}=2$.
}
\end{figure}
\end{multicols}

\begin{references}
\bibitem{NB1}
This estimate uses least-square-fits in the time region
where $\xi_Q$ and $\xi_H$ differ by less than $5\%$.
The data for $\alpha=2$ are taken from long runs up to $t=204800$.
\end{references}
\end{document}